\documentclass[letterpaper, 10 pt, conference]{ieeeconf}

\IEEEoverridecommandlockouts
\usepackage{amsmath,amssymb,amsfonts}
\usepackage{amsthm}
\usepackage{array}
\usepackage{algorithm}
\usepackage{algpseudocode}
\usepackage{graphicx}
\usepackage{textcomp}
\usepackage{xcolor}
\usepackage[hidelinks]{hyperref}
\usepackage{eso-pic}
\usepackage{tikz}
\usetikzlibrary{calc}

\graphicspath{{figures/}}

\theoremstyle{definition}
\newtheorem{theorem}{Theorem}

\theoremstyle{remark}
\newtheorem{remark}[theorem]{Remark}

\definecolor{mplblue}{HTML}{1F77B4}
\definecolor{mplorange}{HTML}{FF7F0E}
\definecolor{mplgreen}{HTML}{2CA02C}
\definecolor{mplred}{HTML}{D62728}

\newcommand{\legsymwidth}{5mm}
\newcommand{\legsymlw}{1.5pt}
\NewDocumentCommand{\legsym}{ m O{solid} O{} }{%
  \tikz[baseline=-0.6ex]{%
    \legsymDrawLine{#1}{#2}%
    \legsymDrawMark{#1}{#3}%
  }%
}
\newcommand{\legsymDrawLine}[2]{%
  \def\legsymTmp{#2}\def\legsymNone{none}%
  \ifx\legsymTmp\legsymNone\else
    \draw[#1,#2,line width=\legsymlw] (0,0) -- (\legsymwidth,0);%
  \fi
}
\newcommand{\legsymDrawMark}[2]{%
  \def\legsymTmp{#2}\def\legsymEmpty{}%
  \ifx\legsymTmp\legsymEmpty\else
    \coordinate (lsm) at ($(0,0)!0.5!(\legsymwidth,0)$);%
    \def\legsymT{circle}\ifx\legsymTmp\legsymT  \filldraw[#1] (lsm) circle[radius=2.1pt];\fi
    \def\legsymT{diamond}\ifx\legsymTmp\legsymT \filldraw[#1] ($(lsm)+(0,2.5pt)$) -- ($(lsm)+(2.5pt,0)$) -- ($(lsm)+(0,-2.5pt)$) -- ($(lsm)+(-2.5pt,0)$) -- cycle;\fi
  \fi
}

\title{\LARGE \bf
Data-Efficient Quadratic Q-Learning Using LMIs
}

\author{J.S. van Hulst, W.P.M.H. Heemels, D.J. Antunes 
\thanks{The research is carried out as part of the ITEA4 20216 ASIMOV project. The ASIMOV activities are supported by the Netherlands Organisation for Applied Scientific Research TNO and the Dutch Ministry of Economic Affairs and Climate (project number: AI211006). The research leading to these results is partially funded by the German Federal Ministry of Education and Research (BMBF) within the project ASIMOV-D under grant agreement No. 01IS21022G [DLR], based on a decision of the German Bundestag.}
\thanks{The authors are with the Control Systems Technology Section, Department of Mechanical Engineering, Eindhoven University of Technology, the Netherlands. Emails:{\tt\small \{j.s.v.hulst, m.heemels, d.antunes\}@tue.nl}}%
}

\AddToShipoutPictureBG*{%
	\AtPageUpperLeft{%
		\setlength\unitlength{1in}%
		\hspace*{\dimexpr0.5\paperwidth\relax}%
		\makebox(0,-1)[c]{\begin{tabular}{>{\centering\arraybackslash}p{\textwidth}}
			J.S. van Hulst \emph{et al.}, Data-Efficient Quadratic Q-Learning Using LMIs, \emph{63rd IEEE Conference on Decision and Control (CDC)}, Milan, Italy, 2024, pp. 1161--1166. DOI:\href{https://doi.org/10.1109/CDC56724.2024.10886653}{10.1109/CDC56724.2024.10886653}. This version contains minor revisions and additional figures compared to the IEEE publication.
		\end{tabular}}
}}

\AddToShipoutPictureBG*{%
\AtPageUpperLeft{%
\setlength\unitlength{1in}%
\hspace*{\dimexpr0.5\paperwidth\relax}
\makebox(0,-21)[c]{
\footnotesize
\begin{tabular}{>{\centering\arraybackslash}p{\textwidth}}
\textcopyright{} 2024 IEEE. Personal use of this material is permitted. Permission from IEEE must be obtained for all other uses, in any current or future media, including reprinting/republishing this material for advertising or promotional purposes, creating new collective works, for resale or redistribution to servers or lists, or reuse of any copyrighted component of this work in other works.
\end{tabular}}}}

\begin{document}

\maketitle
\thispagestyle{empty}
\pagestyle{empty}

\begin{abstract}
Reinforcement learning (RL) has seen significant research and application results but often requires large amounts of training data. This paper proposes two data-efficient off-policy RL methods that use parametrized Q-learning. In these methods, the Q-function is chosen to be linear in the parameters and quadratic in selected basis functions in the state and control deviations from a base policy. A cost penalizing the $\ell_1$-norm of Bellman errors is minimized. We propose two methods: Linear Matrix Inequality Q-Learning (LMI-QL) and its iterative variant (LMI-QLi), which solve the resulting episodic optimization problem through convex optimization. LMI-QL relies on a convex relaxation that yields a semidefinite programming (SDP) problem with linear matrix inequalities (LMIs). LMI-QLi entails solving sequential iterations of an SDP problem. Both methods combine convex optimization with direct Q-function learning to significantly improve learning speed. A numerical case study demonstrates their advantages over least-squares policy iteration, a standard parametrized Q-learning method. Code is available at \url{https://github.com/JvHulst/LMI_Q-Learning}.
\end{abstract}

\section{INTRODUCTION}
Reinforcement learning (RL) has recently received significant attention due to excellent results in many application domains~\cite{Sutton2018}. The data-driven nature of RL methods enables learning of high-performing policies, even without explicit models.

Q-learning, a common RL method, estimates the so-called quality of an action-state pair, representing the expected cumulative reward of choosing an action given a state. This quality learning uses the temporal difference error to update the Q-values iteratively and can be performed online or batch-wise. Q-learning in its most basic form tends to have slow convergence to the optimal policy, due to having to visit every state and action multiple times. Besides, it often requires the discretization of uncountable state and control spaces, which scales poorly as the dimension of the state space increases. To improve convergence speed and circumvent the curse of dimensionality, one can parameterize the Q-function using basis functions~\cite{Busoniu2011,Busoniu2018,Xu2014}.

This paper aims to accelerate Q-function convergence through convex optimization. Our approach consists of two key assumptions. First, the Q-function is chosen to be linear in the parameters and quadratic in a set of selected basis functions in the state, as well as quadratic in the control deviations from a given base policy, as detailed in Section~\ref{sec:method}. Second, we find the Q-function parameters by minimizing the $\ell_1$-norm of the data-based Bellman errors. We propose two numerical methods to tackle the resulting episodic parameter optimization problem by solving convex subproblems. The first method reformulates the Bellman error minimization problem as a semidefinite programming problem by fixing the optimization variables that appear nonlinearly. We can then solve the resulting convex problem, update the previously fixed variables, and repeat the process iteratively. The second numerical method that we propose relies on a \emph{convex relaxation} of the temporal difference minimization problem in the form of a semidefinite program including a set of linear matrix inequalities (LMIs). \emph{Convex relaxations} provide efficient suboptimal solutions for difficult non-convex optimization problems while bounding the original cost function.

Even though both methods rely on convex optimization, the resulting solutions typically differ from the optimal Q-function due to the approximations introduced by coordinate descent and convex relaxations. Nonetheless, practical results demonstrate the effectiveness of the proposed approaches. Importantly, the proposed methods belong to the class of off-policy RL methods, and thus the data used can be generated in any way. We coin the term linear matrix inequality Q-learning (LMI-QL) for the convex relaxation method and linear matrix inequality Q-learning with iterations (LMI-QLi) for the iterative method.

Our proposed methods share similarities with existing approaches like least-squares temporal difference (LSTD) learning~\cite{Bradtke1996}, which uses convex least-squares optimization to improve data efficiency. Extensions such as recurrent least-squares temporal difference (RLSTD) offer further improvements. Least squares policy iteration (LSPI)~\cite{Lagoudakis2004} extends the idea of LSTD to the Q-function. In~\cite{Lagoudakis2014}, an overview of more least-squares RL methods is presented. Among these, LSPI performs approximate policy iteration, alternating policy evaluation with greedy policy improvement. Our methods instead target the optimal Q-function directly by minimizing the Bellman residual, rather than alternating evaluation and improvement. Like our methods, fitted Q-iteration learns the Q-function from a fixed batch of data~\cite{Ernst2005}, including the data-efficient neural variant of Riedmiller~\cite{Riedmiller2005}. The convex structure we add lets each fit attain its global optimum.

Well-known RL methods like DQN~\cite{Mnih2015}, PPO~\cite{Schulman2017}, and MPO~\cite{Abdolmaleki2018} achieve strong performance through various other innovations. DQN employs deep networks, which are extremely flexible but yield slow convergence in large spaces. PPO and MPO focus on stabilizing parameter updates rather than directly improving convergence speed. In contrast, our methods use structured parametrizations and convex optimization to achieve fast learning, particularly in problems where flexibility is less critical.

Although various results in the literature accelerate Q-learning using parametrization structures, to the best of the authors' knowledge, none combine the following two elements
\begin{enumerate}
    \item direct learning of the optimal Q-function, rather than evaluating and improving a sequence of policies;
    \item convex optimization, so that every optimization problem we solve attains its global optimum.
\end{enumerate}
LMI-QL realizes the first element through a single convex program, whereas LMI-QLi solves a short sequence of convex programs. By combining direct Q-function learning with convex optimization, LMI-QL(i) learns accurate policies from little data. A numerical case study demonstrates both methods' effectiveness, comparing LMI-QL(i) to LSPI.

This paper is organized as follows. Section~\ref{sec:problem} presents the problem tackled in this paper, as well as provides some preliminaries on parametric Q-function learning. In Section~\ref{sec:method}, the proposed convex Q-learning methods are presented. In Section~\ref{sec:pendulum_results}, we present the results of applying the methods to a nonlinear pendulum system. Lastly, Section~\ref{sec:conclusions} gives conclusions and recommendations for future research.

\noindent \textbf{Notation.}~Let $\mathbb{R} =(-\infty,\infty)$. Let $\mathbb{N} = \{0,1,\ldots\}$ denote the natural numbers.
The $i$-th element of a vector $x$ is denoted $x[i]$.
The identity matrix of size $n$ is denoted by $I_{n}$.
A normal distribution with mean vector $\mu \in \mathbb{R}^n$ and covariance matrix $\Sigma \in \mathbb{R}^{n \times n}$ is denoted $\mathcal{N}(\mu,\Sigma)$.

\section{PROBLEM FORMULATION}
\label{sec:problem}
The notation in this section is largely taken from~\cite{Sutton2018}. We consider a generic infinite-horizon Markov decision process (MDP) with stochastic state transitions.
We denote the system state at discrete-time index $k \in \mathbb{N}$ by $x_k \in \mathbb{R}^{n_x}$, and the system input (or action) by $u_k \in \mathbb{R}^{n_u}$. The conditioned probability distribution that defines the state transition is fully characterized by $P: \mathbb{R}^{n_x} \times \mathbb{R}^{n_x} \times \mathbb{R}^{n_u} \to [0,\infty)$ where $\text{Prob}\left[x_{k+1}\in A \mid x_k=x,u_k=u\right] = \int_{A} P(x',x,u) dx'$ for $A \subseteq \mathbb{R}^{n_x}$, with $\int_{\mathbb{R}^{n_x}} P(x',x,u) dx'=1$, for every $x\in\mathbb{R}^{n_x}$, $u\in\mathbb{R}^{n_u}$. The reward $r_k \in \mathbb{R}$ immediately following the state and input at time index $k\in \mathbb{N}$ is defined as
\begin{equation}
\label{eq:reward}
    r_k = g(x_k,u_k)
\end{equation}
with $g: \mathbb{R}^{n_x} \times \mathbb{R}^{n_u} \to \mathbb{R}$ the reward function.

The system input is often chosen as a function of the system state according to a so-called policy. We define a deterministic, time-invariant policy $\pi: \mathbb{R}^{n_x} \to \mathbb{R}^{n_u}$ as
\begin{equation}
\label{eq:policy}
    u_k = \pi(x_k), \quad k \in \mathbb{N}
\end{equation}
A trajectory of states and inputs can be constructed, given the state transition dynamics characterized by $P$ and the policy~\eqref{eq:policy}. From such a trajectory, we can obtain the infinite-horizon discounted return, which is given by
\begin{equation}
\label{eq:return}
    G_k = \sum_{i=0}^\infty \gamma^i r_{k+i}
\end{equation}
with $\gamma \in (0,1)$ the discount factor. The central problem in RL is to find a policy $\pi^*$ that maximizes the expected value of the return. This expected value of the return under a policy $\pi$ starting at state $x_k$ is referred to as the value function of that policy, i.e.,
\begin{equation}
\label{eq:vfunction}
    V^\pi(x_k) := \mathbb{E}\left[G_k \mid x_k,\pi\right].
\end{equation}
A related notion is the quality function or Q-function, which is defined as
\begin{equation}
\label{eq:Qfunction}
    Q^\pi(x_k,u_k) := \mathbb{E}\left[g(x_k,u_k) + \gamma V^\pi(x_{k+1})  \mid x_k,u_k,\pi \right].
\end{equation}
The quality function can play a key role in finding the optimal policy $\pi^*$, as
\begin{equation}
\label{eq:optimal_policy}
    \pi^*(x_k) \in \arg\max_u Q^*(x_k,u),
\end{equation}
in which $Q^*$ is the optimal Q-function, defined by
\begin{equation}
\label{eq:optimal_Qfunction}
    Q^*(x_k,u) = \max_\pi Q^\pi(x_k,u).
\end{equation}
In this sense, once the optimal Q-function $Q^*$ is known, the optimal policy $\pi^*$ follows directly via~\eqref{eq:optimal_policy}.

In this paper, we make use of a parametrized quality function $\tilde{Q}(x_k,u_k,\theta)$ with $\theta \in \mathbb{R}^{n_\theta}$ a vector of parameters.
The problem formulation is, given a batch of data $\mathcal{D} := \{\mathcal{X},\mathcal{U},\mathcal{R},\mathcal{X}_+\}$,
\begin{equation*}
\begin{alignedat}{5}
    \mathcal{X} & := \{x_0, ~& x_1, ~& \ldots, ~& x_{N}\},& \\
    \mathcal{U} & := \{u_0, ~& u_1, ~& \ldots, ~& u_{N}\},& \\
    \mathcal{R} & := \{r_0, ~& r_1, ~& \ldots, ~& r_{N}\},&\quad \text{and} \\
    \mathcal{X}_+ & := \{x_1, ~& x_2, ~& \ldots, ~& x_{N+1}\},&
\end{alignedat}
\end{equation*}
find the parameters $\theta$ that yield Q-function $\tilde{Q}$, which best fits the data (in an approximated sense) according to the definition of the optimal quality function~\eqref{eq:optimal_Qfunction}.  In this sense, we are minimizing the ``distance'' between the parametrized Q-function $\tilde{Q}$ and the optimal Q-function $Q^*$, given the data. Note that the dataset $\mathcal{D}$ should be sufficiently informative, i.e., persistently exciting (PE) the parametrized quality function. We will later consider how this is achieved for the specific quality function structure proposed in this paper.

\section{PROPOSED METHOD}
\label{sec:method}
In this section, the proposed method is presented. We start by formalizing the problem setup by presenting the proposed cost and Q-function parametrization. Note that the cost discussed in this section refers to the objective function associated with the Q-function parameter optimization problem, rather than the MDP reward function. Afterward, we detail two proposed numerical methods.

\subsection{Formalization of the Problem Setup}
We propose to capture the objective of the previous section in the optimization problem
\begin{equation}
\label{eq:norm_cost}
    \min_\theta  \left\lVert z \right\rVert_p
\end{equation}
with $z := \begin{bmatrix}  z_0, z_1, \ldots, z_N  \end{bmatrix}^\top$, in which
\begin{equation*}
    z_k := \tilde{Q}(x_k,u_k,\theta) - \left(r_k + \gamma \max_{u} \tilde{Q}(x_{k+1},u,\theta)\right),
\end{equation*}
for $k \in \{0, 1, \ldots, N\}$.
Here, $\lVert x \rVert_p$ represents the $\ell_p$-norm of $x$. The quantity $z_k$ is often referred to as the temporal difference or Bellman residual. This optimization problem captures the central problem of parametric Q-learning since the Bellman residuals are zero for the optimal Q-function. The size of the residual also relates to the distance to $Q^*$. For the exact Bellman residual, if $\sup_{x,u} \lvert \tilde{Q}(x,u) - (\mathcal{T}\tilde{Q})(x,u) \rvert \leq \epsilon$ with $\mathcal{T}$ the Bellman optimality operator, then $\lVert \tilde{Q} - Q^* \rVert_\infty \leq \epsilon / (1-\gamma)$~\cite{Bertsekas1996}. This bound is idealized here, since it assumes a uniform bound over the whole state-action space, while we minimize the residual only on the finite sample $\mathcal{D}$. Minimizing the residual directly, as we do here, also avoids the divergence that semi-gradient temporal-difference methods can exhibit under off-policy data and function approximation~\cite{Baird1995,Sutton2018}. Note that in LSPI~\cite{Lagoudakis2004}, the $\ell_2$-version of this problem is considered. In this paper, we will use the $\ell_1$-norm ($p=1$) or absolute value norm. This particular choice is made since an absolute value norm cost admits a (convex) semidefinite programming form, as we will see.

It is assumed we have access to a baseline policy along with an approximation of its value function. The baseline policy is denoted by $\bar{\pi}: \mathbb{R}^{n_x} \to \mathbb{R}^{n_u}$, while the corresponding value function is denoted by $V^{\bar{\pi}}: \mathbb{R}^{n_x} \to \mathbb{R}$. We intend to learn the deviation from this baseline, i.e., $\Delta: \mathbb{R}^{n_x} \to \mathbb{R}^{n_u}$ in
\begin{equation}
    u_k = \pi(x_k) = \bar{\pi}(x_k) + \Delta(x_k).
\end{equation}
If no base policy is available, we can assume $\bar{\pi} = V^{\bar{\pi}} = 0$ and model the full Q-function with the parametrization that follows.

\begin{figure*}[t!]
  \centering
    \includegraphics[width=0.72\textwidth]{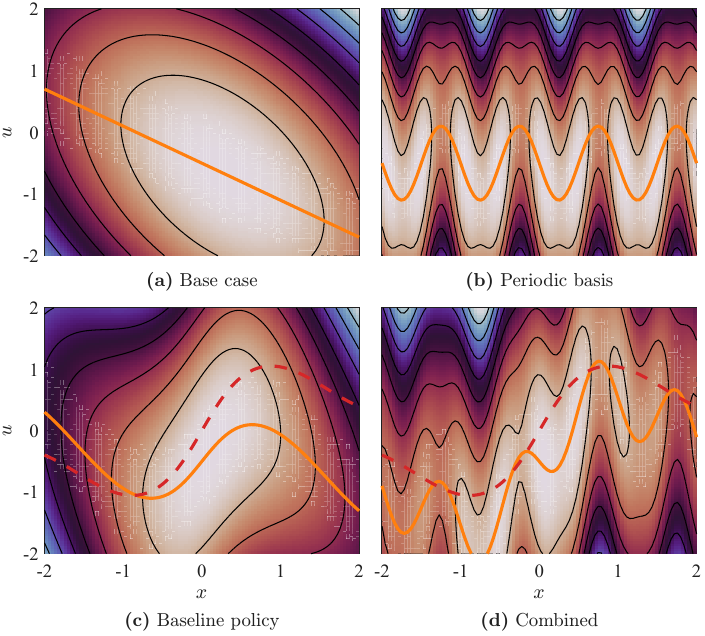}
  \caption{%
    Quadratic Q-function parametrization~\eqref{eq:quadratic_Qfunction} for a scalar state $x$ and scalar input $u$, shown over $x,u\in[-2,2]$. In every panel the color indicates the value of the parametrized Q-function $\tilde{Q}(x,u)$ and the black curves are its level sets. Shown are the greedy policy $\tilde{\pi}(x)$ (\protect\legsym{mplorange}) from~\eqref{eq:quadratic_Qfunction_greedy_action}, and the baseline policy $\bar\pi(x)$ (\protect\legsym{mplred}[dashed]).
    The simplest case~\textbf{(a)} has identity basis $\phi(x)=x$, zero baseline policy $\bar\pi(x)=0$ and zero baseline value $V^{\bar{\pi}}(x)=0$; a nonzero linear term $R=[0;\,-1]$ places the maximizer at $\tilde{\pi}(0)=-0.5$, so the policy does not pass through the origin.
    The same case with a periodic basis $\phi(x)=\sin(2\pi x)$~\textbf{(b)} makes the level sets repeat along $x$.
    A baseline policy $\bar\pi(x)=0.7\tanh(2x)+0.4\sin(2x)$ together with a baseline value $V^{\bar{\pi}}(x)=-0.7x^{2}+0.5\cos(2x)$ and identity basis $\phi(x)=x$~\textbf{(c)} bends the ridge of maximizers, and the baseline value is added to the Q-function, reshaping it along $x$.
    The periodic basis, the baseline policy and the baseline value are combined in~\textbf{(d)}.}
  \label{fig:qfunction}
\end{figure*}

Given the policy baseline, we propose to parametrize the Q-function as
\begin{equation}
\label{eq:quadratic_Qfunction}
\begin{aligned}
    \tilde{Q}(x,u,\theta) =&~ V^{\bar{\pi}}(x)+T+\begin{bmatrix}
        \phi(x) \\ u-\bar{\pi}(x)
    \end{bmatrix}^\top R \\
    & \quad - \begin{bmatrix}
        \phi(x) \\ u-\bar{\pi}(x)
    \end{bmatrix}^\top S \begin{bmatrix}
        \phi(x) \\
        (u-\bar{\pi}(x))
    \end{bmatrix},
\end{aligned}
\end{equation}
where $\phi: \mathbb{R}^{n_x} \to \mathbb{R}^{n_{\phi}}$ is a vector of basis functions in the state variable $x$ and in which $T \in \mathbb{R}$, $R := \begin{bmatrix}
    R_x \\ R_u
\end{bmatrix} \in \mathbb{R}^{n_{\phi}+n_u}$, and $S := \begin{bmatrix}
    S_{xx} & S_{xu} \\ S_{xu}^\top & S_{uu}
\end{bmatrix} \in \mathbb{R}^{(n_{\phi}+n_u) \times (n_{\phi}+n_u)}$ with $S=S^\top \succ 0$. The parameter vector $\theta$ contains all the elements of $T$, $R$, and the upper-triangular elements of $S$. Hence, $n_\theta = 1 + n_{\phi}+n_u+\frac{(n_{\phi}+n_u)(n_{\phi}+n_u+1)}{2}$. With the stacked vector $\zeta := \begin{bmatrix} \phi(x)^\top & (u-\bar{\pi}(x))^\top \end{bmatrix}^\top$, the parametrization~\eqref{eq:quadratic_Qfunction} reads $\tilde{Q}(x,u,\theta) = V^{\bar{\pi}}(x) + T + \zeta^\top R - \zeta^\top S \zeta$. Fig.~\ref{fig:qfunction} provides intuition behind the proposed parametrization.

Completing the square in~\eqref{eq:quadratic_Qfunction} gives
\begin{equation}
\label{eq:quadratic_Qfunction_completed_square}
\begin{aligned}
    \tilde{Q}(x,u,\theta) =&~ V^{\bar{\pi}}(x) + T + \tfrac{1}{4}R^\top S^{-1} R \\
    & - \left( \zeta - \tfrac{1}{2}S^{-1}R \right)^\top S \left( \zeta - \tfrac{1}{2}S^{-1}R \right).
\end{aligned}
\end{equation}
Since $S \succ 0$, the Q-function is strictly concave in $\zeta$, with unconstrained maximizer $\zeta = \tfrac{1}{2}S^{-1}R$, so the linear term $R$ sets the location of this maximizer. In particular, $S_{uu} \succ 0$ makes $\tilde{Q}$ strictly concave in $u$ for every state $x$, so the greedy action is unique, as we derive next. If $\phi(x)=x$ (and the baseline $\bar{\pi}$ is affine), then $\zeta$ is affine in $(x,u)$, and the Q-function is strictly concave in $(x,u)$ as well whenever the baseline value $V^{\bar{\pi}}$ is concave. For a nonlinear basis $\phi$, the map $x \mapsto \phi(x)$ is generally not invertible, so concavity in $\zeta$ does not imply a single peak in the original variables $(x,u)$. In the LQR setting with linear dynamics and quadratic reward, the choice $\phi(x)=x$ makes the quality function exactly quadratic in $(x,u)$.

Given a Q-function parametrization, the corresponding greedy action can be found by setting the partial derivatives with respect to $u$ equal to zero and solving for $u$. For the particular parametrization in~\eqref{eq:quadratic_Qfunction}, we obtain
\begin{equation}
\label{eq:quadratic_Qfunction_greedy_action}
\begin{aligned}
    \tilde{\pi}(x) =&~ \arg \max_u \tilde{Q}(x,u,\theta) \\
    =&~  \bar{\pi}(x)+\frac{1}{2}S_{uu}^{-1}R_u-S_{uu}^{-1}S_{xu}^\top \phi(x).
\end{aligned}
\end{equation}
Note that the basis functions $\phi$ appear linearly in this greedy policy. In this sense, $\phi$ allows us to choose an affine feedback parametrization. The choice of $\phi$ also informs the PE requirements on the dataset $\mathcal{D}$, which is explained in the following remark.
\begin{remark}
    Even though the proposed methods are off-policy, the data-generating policy cannot be selected arbitrarily. We need stochasticity in the policy during the creation of dataset $\mathcal{D}$, such that the inputs persistently excite the basis functions $\phi$ through the dynamics. In this case, the PE condition is satisfied if the parameter optimization problem~\eqref{eq:norm_cost} with~\eqref{eq:quadratic_Qfunction} has a unique solution. The basis $\phi$ is chosen by hand. Its choice fixes both the achievable policy class, through the greedy action~\eqref{eq:quadratic_Qfunction_greedy_action}, and the persistence-of-excitation requirement on $\mathcal{D}$.
\end{remark}

When we substitute~\eqref{eq:quadratic_Qfunction_greedy_action} into the parametrized Q-function~\eqref{eq:quadratic_Qfunction}, we obtain
\begin{equation}
\label{eq:quadratic_Qfunction_greedy_Qvalue}
\small
    \tilde{Q}(x,\tilde{\pi}(x),\theta) =  V^{\bar{\pi}}(x) + T + \begin{bmatrix}
        \phi(x) \\ 1
    \end{bmatrix}^\top \left(\Omega + \Psi S_{uu}^{-1} \Psi^\top \right) \begin{bmatrix}
        \phi(x) \\ 1
    \end{bmatrix}
\end{equation}
with
{\small
\begin{equation*}
    \Omega := \begin{bmatrix}
    -S_{xx} & \frac{1}{2}R_x \\  \frac{1}{2}R_x^\top & 0
\end{bmatrix}, \quad \Psi := \begin{bmatrix}
    S_{xu} \\ -\frac{1}{2}R_u^\top
\end{bmatrix},
\end{equation*}}%
which presents the Q-value that follows from the so-called greedy action. With the chosen cost function and Q-function structure, we can move to solutions to the parameter optimization problem. The next section details a convex numerical method to this effect.

\subsection{Iterative Convex Method}
Optimization problems with $\ell_1$-norm cost can be equivalently characterized with linear cost and inequalities containing the original cost vector. If we perform this transformation and substitute~\eqref{eq:quadratic_Qfunction} and~\eqref{eq:quadratic_Qfunction_greedy_Qvalue} into~\eqref{eq:norm_cost} with $p=1$, we obtain
\begin{equation}
\label{eq:quadratic_Qfunction_cost}
\begin{aligned}
    \min_{\theta} ~~& \sum_{k=0}^N  t_k,  \\
    \text{s.t.} ~~& t_k \geq \tilde{z}_k, \quad &  k \in \{0,1,\ldots, N\},\\
    ~~& t_k \geq - \left(\tilde{z}_k\right), \quad &  k \in \{0,1,\ldots, N\},\\
    ~~& W = \Psi S_{uu}^{-1} \Psi^\top,\\
    ~~& \Omega = \begin{bmatrix}
    -S_{xx} & \frac{1}{2}R_x \\  \frac{1}{2}R_x^\top & 0
\end{bmatrix}, \\
    ~~& S\succ 0,
\end{aligned}
\end{equation}
where $t_k \in \mathbb{R}$ for $k \in \{0,1,\ldots, N\}$ and $\Omega, W \in \mathbb{R}^{(n_\phi + 1) \times (n_\phi + 1)}$ are auxiliary variables, and in which
{\small
\begin{equation*}
\begin{aligned}
    \tilde{z}_k :=&~ \tilde{Q}(x_k,u_k,\theta) - \Biggl(r_k + \gamma \biggl(V^{\bar{\pi}}(x_{k+1}) + T\\
    & \quad\quad + \begin{bmatrix}
        \phi(x_{k+1}) \\ 1
    \end{bmatrix}^\top \left(\Omega+W\right) \begin{bmatrix}
        \phi(x_{k+1}) \\ 1
    \end{bmatrix}\biggr)\Biggr).
\end{aligned}
\end{equation*}}%
This optimization problem has a cost function that is linear in the design variables and linear matrix inequality constraints, along with a single nonlinear matrix equality constraint $W = \Psi S_{uu}^{-1} \Psi^\top$.

To make the optimization problem convex, we fix the optimization variables that appear nonlinearly in $\eqref{eq:quadratic_Qfunction_cost}$ by setting $W$ to arbitrary feasible values $\hat{\Psi}$, $\hat{S}_{uu}$. Additionally, we propose fixing the variable $\Omega$ to feasible values $\hat{S}_{xx}$ and $\hat{R}_x$, even though $\Omega$ appears linearly in the optimization problem. While this step is not required for convexity, empirical results suggest it often leads to improved solutions. After these adjustments, the optimization problem becomes
\begin{equation}
\label{eq:policy_iteration_cost}
    \begin{aligned}
        \min_{\theta} ~~& \sum_{k=0}^N  t_k,  \\
        \text{s.t.} ~~& t_k \geq \tilde{z}_k, \quad &  k \in \{0,1,\ldots, N\},\\
        ~~& t_k \geq - \left(\tilde{z}_k\right), \quad &  k \in \{0,1,\ldots, N\},\\
        ~~& W = \hat{\Psi} \hat{S}_{uu}^{-1} \hat{\Psi}^\top,\\
        ~~& \Omega = \begin{bmatrix}
    -\hat{S}_{xx} & \frac{1}{2}\hat{R}_x \\  \frac{1}{2}\hat{R}_x^\top & 0
\end{bmatrix}, \\
        ~~& S\succ 0.
    \end{aligned}
\end{equation}
This optimization problem is convex, so any local minimizer is also a global minimizer. In fact, since the cost is linear in the design variables and the constraints are LMIs, the optimization problem~\eqref{eq:policy_iteration_cost} is a semidefinite programming problem, for which a wide variety of efficient numerical solvers are available. Note that in numerical implementations, e.g., using SDP solvers, strict inequalities are often replaced by non-strict ones. After solving it, we can update $\hat{S}_{xx}$, $\hat{R}_x$, $\hat{\Psi}$ and $\hat{S}_{uu}$ using the solution $\theta^*$ and repeat the optimization process. This implementation resembles coordinate descent since we optimize over a subset of design variables. Fixing the next-state value before refitting the Q-function also makes each iteration closely related to fitted Q-iteration~\cite{Ernst2005}, for which error-propagation results relate the closed-loop performance of the resulting policy to the per-iteration Bellman error~\cite{Bertsekas1996}. These results assume the true Bellman error. For stochastic transitions, the single-sample residual used here is a biased estimate of it~\cite{Baird1995}.

At a fixed point of this iteration, the updated values equal those of the solution, that is, $\hat{S}_{xx}=S_{xx}$, $\hat{R}_x=R_x$, $\hat{\Psi}=\Psi$ and $\hat{S}_{uu}=S_{uu}$. The equality constraint $W=\Psi S_{uu}^{-1}\Psi^\top$ of~\eqref{eq:quadratic_Qfunction_cost} then holds, so the iterate is feasible for the original problem. It need not minimize the original problem, since each solve holds the next-state value fixed and ignores how that value varies with $\theta$. This behavior mirrors fixed-target iterations in system identification, such as the Sanathanan--Koerner iteration~\cite{Sanathanan1963}, whose fixed point likewise need not be a stationary point of the original problem. There are no guarantees that a fixed point is reached after finitely many iterations, yet we observe good results in practice.

The next section details an alternative solution to the optimization problem~\eqref{eq:quadratic_Qfunction_cost} using a convex relaxation.

\subsection{Convex Relaxation}
Consider again the optimization problem in~\eqref{eq:norm_cost} with $p=1$ and $\tilde{Q}$ parametrized according to~\eqref{eq:quadratic_Qfunction}. A \emph{convex relaxation} of this problem is given by the optimization problem
\begin{equation}
\label{eq:relaxed_cost}
\begin{aligned}
    \min_{\theta} ~~& \sum_{k=0}^N  t_k  \\
    \text{s.t.} ~~& t_k \geq \tilde{z}_k, \quad & k \in \{0,1,\ldots, N\},\\
    ~~& t_k \geq - \left(\tilde{z}_k\right), \quad & k \in \{0,1,\ldots, N\},\\
    ~~& \begin{bmatrix}
        W & \Psi \\
        \Psi^\top & S_{uu}
    \end{bmatrix}\succeq 0,\\
    ~~& \Omega = \begin{bmatrix}
    -S_{xx} & \frac{1}{2}R_x \\  \frac{1}{2}R_x^\top & 0
\end{bmatrix}, \\
    ~~& S\succ 0.
\end{aligned}
\end{equation}
We obtain this relaxation by replacing the equality constraint $W=\Psi S_{uu}^{-1} \Psi^\top$ by the inequality constraint $W \succeq \Psi S_{uu}^{-1} \Psi^\top$. Afterward, since $S \succ 0$ implies that $S_{uu} \succ 0$, we can apply the Schur complement. We obtain the equivalent matrix inequality included as a constraint in~\eqref{eq:relaxed_cost} which is linear in the design parameters. The resulting \emph{relaxed} optimization problem~\eqref{eq:relaxed_cost} is a (convex) semidefinite programming (SDP) problem, as the cost is linear in the design variables and the constraints are LMIs. Furthermore, it is a relaxation of~\eqref{eq:quadratic_Qfunction_cost} since the cost function is unchanged and the feasible domain is (slightly) expanded.

The Schur complement step above uses only $S_{uu} \succ 0$, so the full constraint $S \succ 0$ is not required for convexity of the reformulation, nor for the uniqueness of the greedy action~\eqref{eq:quadratic_Qfunction_greedy_action}. We retain $S \succ 0$ as a modeling choice, since it makes the parametrized Q-function strictly concave in the stacked vector $\zeta$.

The convex relaxation provides guaranteed lower and upper bounds on the optimal cost. The lower bound arises because the relaxed optimal cost can never exceed the original, as the relaxation expands the feasible set. Moreover, the relaxed solution for $\theta$ is feasible for the original problem~\eqref{eq:quadratic_Qfunction_cost} if we set $W = \Psi S_{uu}^{-1} \Psi^\top$, allowing us to evaluate an upper bound. If the bounds are tight, then the relaxed solution is also optimal for the original problem.

In order to make the relaxation less conservative, we can add the matrix inequality
\begin{equation}
\label{eq:extra_matrix_inequality}
    S_{xx} \succ W_{11},
\end{equation}
where $W_{11} \in \mathbb{R}^{n_{\phi} \times n_{\phi}}$ is the top left submatrix of $W$. This matrix inequality follows from the original non-relaxed optimization problem by taking the Schur complement of $S$. The introduction of the matrix inequality~\eqref{eq:extra_matrix_inequality} constrains $W$ from above, potentially decreasing the gap between $W$ and $\Psi S_{uu}^{-1} \Psi^\top$.

Next, we introduce a penalty term to the cost function, such that the inequality $W\succeq \Psi S_{uu}^{-1} \Psi^\top$ becomes closer to equality. This penalty term can be chosen in many different ways. In this paper, we choose the nuclear norm of $W$, i.e.,
\begin{equation}
    \lVert W \rVert_* := \sum_{i=1}^{n_{\phi}+1} \sigma_i
\end{equation}
where $\sigma_i$ are the singular values of $W$. Note that since $W$ is square and positive semidefinite, $\lVert W \rVert_* = \text{tr}(W)$.

The choice of the nuclear norm as the penalty function is motivated by several factors. First, the trace operator maintains the convexity of the optimization problem. Second, penalizing the nuclear norm --- equal to the sum of the eigenvalues of $W$ --- directly addresses the gap between $W$ and $\Psi S_{uu}^{-1} \Psi^\top$ by discouraging excessively large eigenvalues of $W$. Last, in typical applications we see that $n_u<n_{\phi}+1$, which results in $\Psi S_{uu}^{-1} \Psi^\top$ being rank-deficient. The nuclear norm presents a convexification of matrix rank minimization~\cite{Recht2010} and can therefore be used to promote rank-deficient $W$. This combination of a Schur-complement relaxation with a trace penalty resembles the cone-complementarity linearization of El Ghaoui \emph{et al.}~\cite{ElGhaoui1997}. Conditions under which such nuclear-norm penalization recovers the exact low-rank solution, and hence closes the relaxation gap, are known in special cases~\cite{Recht2010}.

Together with the previous idea, this leads to the following optimization problem
\begin{equation}
\label{eq:relaxed_cost_with_penalty}
\begin{aligned}
    \min_{\theta} ~~& \sum_{k=0}^N  t_k + \lambda ~ \text{tr}(W),  \\
    \text{s.t.} ~~& t_k \geq \tilde{z}_k, \quad & k \in \{0,1,\ldots, N\},\\
    ~~& t_k \geq - \left(\tilde{z}_k\right), \quad & k \in \{0,1,\ldots, N\},\\
    ~~& \begin{bmatrix}
        W & \Psi \\
        \Psi^\top & S_{uu}
    \end{bmatrix}\succeq 0,\\
    ~~& \Omega = \begin{bmatrix}
    -S_{xx} & \frac{1}{2}R_x \\  \frac{1}{2}R_x^\top & 0
\end{bmatrix}, \\
    ~~& S\succ 0,\\
    ~~& S_{xx}\succ W_{11}.
\end{aligned}
\end{equation}

Given a dataset $\mathcal{D}$, the penalty weight $\lambda$ can be chosen by line search to minimize the original cost~\eqref{eq:quadratic_Qfunction_cost}. Section~\ref{subsec:lambda} demonstrates this for the pendulum.

\subsection{Algorithms}
Combining all the ideas presented, we can summarize the LMI-based parametrized Q-learning methods in two pseudocode algorithms. \textbf{Algorithm~\ref{alg:LMI-QL}} details the LMI-based Q-learning method that uses convex relaxation (LMI-QL), while
\textbf{Algorithm~\ref{alg:LMI-QLi}} details the iterative LMI-based Q-learning method that resembles coordinate descent (LMI-QLi), with $\tau \in \mathbb{N}$ the number of iterations.

\begin{algorithm}[ht]
\caption{LMI-QL}\label{alg:LMI-QL}
\begin{algorithmic}[1]
\State Choose $\bar{\pi}$ along with an (approximate) value function $V^{\bar{\pi}}$
    \State Generate $\mathcal{D}$ using an arbitrary (stochastic) policy
    \Repeat
        \State Choose $\lambda$ using line search
        \State Find $\theta^*$, the minimizing solution to~\eqref{eq:relaxed_cost_with_penalty}
    \Until the line search stopping condition is reached
\end{algorithmic}
\end{algorithm}

\begin{algorithm}[ht]
\caption{LMI-QLi}\label{alg:LMI-QLi}
\begin{algorithmic}[1]
\State Choose $\bar{\pi}$ along with an (approximate) value function $V^{\bar{\pi}}$
\State Choose $\tau \in \mathbb{N}$
    \State Generate $\mathcal{D}$ using an arbitrary (stochastic) policy
    \State Initialize $\hat{S}_{xx}$, $\hat{R}_x$, $\hat{\Psi}$, $\hat{S}_{uu}$
    \Repeat
        \State Find $\theta^*$, the minimizing solution to~\eqref{eq:policy_iteration_cost}
        \State Update $\hat{S}_{xx}$, $\hat{R}_x$, $\hat{\Psi}$, $\hat{S}_{uu}$ using $\theta^*$
        \State $\tau \gets \tau-1$
    \Until $\tau=0$
\end{algorithmic}
\end{algorithm}

The two methods can be used to complement each other. We can use LMI-QL to obtain a lower bound on the optimal $\ell_1$ cost, and then use the solution to warm-start LMI-QLi. In terms of computational cost, both methods reduce to semidefinite programs whose size grows linearly in the dataset size, with $O(N)$ variables and constraints. For a fixed $\lambda$, LMI-QL solves one such program, whereas LMI-QLi solves $\tau$ of them, and is therefore a factor $\tau$ more expensive. The line search over $\lambda$ multiplies the cost of LMI-QL by the number of values tried.

In the next section, we will demonstrate the merit of the proposed method in a simulation case study.

\section{PENDULUM SYSTEM SIMULATION CASE STUDY}
\label{sec:pendulum_results}
We apply the proposed methods to a nonlinear pendulum and learn the deviation from a given baseline policy. The goal is to stabilize the pendulum in the upright position.

\subsection{Simulated Setup}
The pendulum has mass $m$ and length $l$ and is affected by gravity with gravitation constant $g$. Furthermore, the pendulum is subject to linear damping proportional to the angular velocity with damping constant $d$. The pendulum equations are discretized in time using the forward Euler method with a sample time of $T_s=0.01$. The state representation of the pendulum consists of its angle and angular velocity. The dynamical model after discretization is given by
{\small
\begin{equation}
\label{eq:pendulum_system}
    x_{k+1} = \begin{bmatrix}
        \text{mod}\left( x_{k}[1] + T_s x_{k}[2] + \pi, 2\pi\right)-\pi \\
        x_{k}[2] + \frac{T_s g}{l}\sin{(x_{k}[1])}-\frac{T_s d}{ml^2}x_{k}[2] + \frac{T_s}{ml^2} u_k
    \end{bmatrix}
    + w_k,
\end{equation}}%
where $m=0.1, l=1, g=9.81, d=0.1$, and with $w_k$ as an independent and identically distributed (i.i.d.) zero-mean Gaussian disturbance with covariance matrix $\Sigma_w = 2.5 \cdot 10^{-5} I_{n_x}$. The reward function is quadratic, i.e., \begin{equation}
\label{eq:quadratic_reward}
    r_k = -\begin{bmatrix}
        x_k^\top & u_k^\top
    \end{bmatrix} M\begin{bmatrix}
        x_k \\
        u_k
    \end{bmatrix},
\end{equation}
in which $M=M^\top \succ 0$, here $M = \text{diag}(1,10^{-2},10^{-5})$. The discount factor in~\eqref{eq:return} is given by $\gamma=0.98$. The states in $\mathcal{D}$ are drawn i.i.d. from $\mathcal{N}(0,(\pi/2)^2 I_{n_x})$. We use an exploring policy, namely, $u_k \sim \mathcal{N}(0,10^1 I_{n_u})$ saturated to $[-5,5]$ at every timestep $k$. The state values in the dataset $\mathcal{D}$ are subjected to i.i.d. zero-mean Gaussian noise with covariance matrix $\Sigma_v = 10^{-5} I_{n_x}$. We choose
\begin{equation*}
    \phi(x) = \begin{bmatrix}
        x \\
        \sin(x[1])
    \end{bmatrix},
\end{equation*}
which allows our controller to counteract the nonlinear effects of gravity on the pendulum. The baseline controller $\bar{\pi}$ is an LQR state feedback controller based on a linearization of~\eqref{eq:pendulum_system} around the origin, with inaccurate parameters $\bar{m} = 0.1 m$ and $\bar{l}=0.3 l$. We set $R=0$, which by~\eqref{eq:quadratic_Qfunction_completed_square} places the maximizer of the Q-function at $\zeta=0$, that is, at the origin $x=0$ with input $u=\bar{\pi}(0)=0$. This is where the reward~\eqref{eq:quadratic_reward} is maximized in this specific case study.

The baseline value function is selected as $V^{\bar{\pi}}(x) = -x^\top \bar{P} x$, with $\bar{P}$ the solution to the algebraic Riccati equation for the inaccurate linearized model. We compare the proposed method to a model-based controller consisting of feedback linearization and LQR on the resulting linear model (with accurate model parameters). This controller is a strong model-based baseline, as it uses the accurate model parameters and enjoys an inverse-optimality property~\cite{Freeman1996}. We also compare the proposed method to the learning method least-squares policy iteration (LSPI), which is detailed in the next section. All reported $\ell_1$ costs are the sums in~\eqref{eq:quadratic_Qfunction_cost} divided by the number of samples, so they are mean absolute Bellman residuals, and $\lambda$ is relative to that scaling.

\subsection{Least-Squares Policy Iteration}
\label{subsec:LSPI}
In LSPI, we perform policy evaluation steps called LSTD-Q, followed by policy improvement steps based on the updated Q-function. The Q-function parametrization in LSPI is linear in the parameters, i.e.,
\begin{equation}
\label{eq:lin_Qfunction}
    \tilde{Q}^{LSPI}(x,u,\theta) = \psi^\top(x,u) \theta^{LSPI},
\end{equation}
with $\theta^{LSPI} \in \mathbb{R}^{n_\theta}$, and $\psi(x,u) := \begin{bmatrix}
    \psi_1(x,u), & \psi_2(x,u), & \ldots, & \psi_{n_\theta}(x,u)
\end{bmatrix}^\top$ a vector of basis functions, with $\psi_i: \mathbb{R}^{n_x} \times \mathbb{R}^{n_u} \to \mathbb{R}$, $i \in \{1,2,\ldots, n_\theta\}$. Note that this Q-function parametrization slightly generalizes~\eqref{eq:quadratic_Qfunction}.
The linearity of $\tilde{Q}^{LSPI}$ in $\theta^{LSPI}$ allows us to reformulate the policy evaluation step as a convex least-squares problem with a closed-form solution by optimizing the parameters using~\eqref{eq:norm_cost} with $p=2$. Importantly, this requires that we fix the policy at the next timestep. In the simulated case study, we use the same Q-function parametrization for LSPI as in the proposed method, i.e., \eqref{eq:quadratic_Qfunction}. Note that LSPI cannot include a positive definite constraint on $S$.

\subsection{Tuning the Penalty \texorpdfstring{$\lambda$}{lambda}}
\label{subsec:lambda}
\begin{figure}[t!]
    \centering
    \includegraphics[width=\columnwidth]{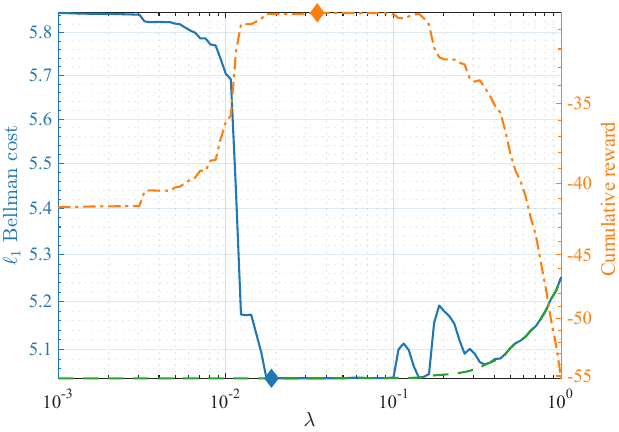}
    \caption{Effect of the penalty weight $\lambda$ on the pendulum dataset. The left axis shows the upper bound (\protect\legsym{mplblue}, original cost at the relaxed solution) and lower bound (\protect\legsym{mplgreen}[dashed], relaxed cost) on the optimal $\ell_1$ Bellman cost, and the right axis shows the closed-loop cumulative reward (\protect\legsym{mplorange}[dash dot]). All axes use a logarithmic scale. The blue marker (\protect\legsym{mplblue}[none][diamond]) indicates the $\lambda$ that minimizes the upper bound and the orange marker (\protect\legsym{mplorange}[none][diamond]) the $\lambda$ that maximizes the reward, which align closely.}
    \label{fig:lambda_search_pendulum}
\end{figure}

We choose the penalty weight $\lambda$ in~\eqref{eq:relaxed_cost_with_penalty} by line search. For each $\lambda$, we solve the relaxed problem and evaluate the original cost~\eqref{eq:quadratic_Qfunction_cost} at the relaxed solution, and we keep the $\lambda$ that minimizes this original cost. Fig.~\ref{fig:lambda_search_pendulum} shows the lower and upper bounds on the optimal $\ell_1$ cost as functions of $\lambda$ for the pendulum dataset, together with the closed-loop reward. The line search is performed offline. Automating the choice of $\lambda$ is a direction for future work.

\subsection{Results}
\begin{figure}[t!]
    \centering
    \includegraphics[width=\columnwidth]{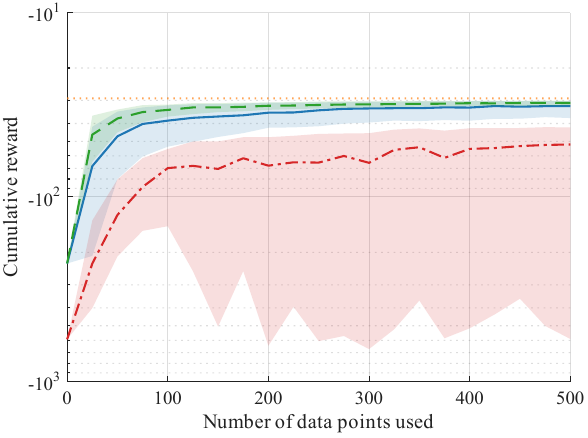}
    \caption{Median cumulative reward and 25th--75th percentile band over 500 Monte Carlo runs of a 100-sample simulation, against the number of data points used, with the reward on a logarithmic axis. Runs with cumulative reward below $-2000$ are excluded as unstable. We compare LQR with feedback linearization (\protect\legsym{mplorange}[dotted]), LMI-QL (\protect\legsym{mplblue}), LMI-QLi (\protect\legsym{mplgreen}[dashed]) and LSPI (\protect\legsym{mplred}[dash dot]).}
    \label{fig:method_comparison_pendulum}
\end{figure}

Fig.~\ref{fig:method_comparison_pendulum} shows the results of applying LSPI, feedback linearization + LQR, and LMI-QL(i) to the pendulum example. We construct a dataset $\mathcal{D}$ with 500 randomly generated initial states, stochastic inputs, rewards, and next states. The learning methods then learn from zero using progressively larger subsets of $\mathcal{D}$ to assess learning speed. For each subset, we run 20 LSPI iterations and 10 LMI-QLi iterations. The methods are compared based on cumulative reward over a 100-sample simulation with identical initial conditions and disturbances. To account for variability, we perform 500 Monte Carlo runs. Evaluations whose cumulative reward falls below $-2000$ are treated as unstable and excluded before computing the median and percentile band, which affects $20.6\%$ of the LSPI evaluations but only $0.83\%$ of the LMI-QL evaluations and $0.27\%$ of the LMI-QLi evaluations.

We also assess the relaxation gap on the full dataset of $N=500$ samples. Taking the median over the 500 seeds at a fixed penalty weight $\lambda=0.02$, LMI-QL attains a relaxed cost of $3.602$, a lower bound on the optimal $\ell_1$ cost, and an original cost of $4.042$ at the relaxed solution, an upper bound. The median of the per-run relative gap $(\text{upper}-\text{lower})/\text{upper}$ is $10.3\%$. The gap is dataset-dependent: about a third of the Monte Carlo runs are nearly tight, with a gap below $1\%$, while around a fifth exceed $20\%$. The near-coincident bounds in Fig.~\ref{fig:lambda_search_pendulum} reflect a dataset on which $\lambda$ was directly tuned, which results in a smaller gap. LMI-QLi attains a median original cost of $3.777$, which presents a tighter upper bound than LMI-QL.

In Fig.~\ref{fig:method_comparison_pendulum}, we can observe that both proposed methods converge close to the cumulative reward that feedback linearization + LQR can achieve. At zero data used, LMI-QL and LMI-QLi start from the baseline policy $\bar\pi$, while LSPI starts from the zero-input policy. This baseline head start lets the proposed methods begin ahead of LSPI before any data has been used, since the baseline provides a good starting point for the Q-function values. Lastly, observe that the proposed methods converge faster than LSPI. This can be explained by the fact that the LSPI solution does not satisfy the $S\succ0$ constraint at multiple points in Fig.~\ref{fig:method_comparison_pendulum}, and by the fact that the proposed methods more directly learn the optimal parametrized Q-function.

\section{CONCLUSIONS}
\label{sec:conclusions}
This paper introduces LMI-QL and LMI-QLi, two methods for direct optimal Q-function learning using convex optimization. We show that if the Q-function is parametrized in a quadratic form and if the objective function uses the $\ell_1$-norm of the Bellman residuals, we can derive two separate convex optimization problems. In LMI-QLi, we fix a subset of the design variables such that the optimization problem for the remaining variables becomes a semidefinite program. In LMI-QL, we relax the temporal difference minimization problem into a semi-definite program including a set of linear matrix inequalities. Simulated results indicate high data efficiency in comparison to least-squares policy iteration for a nonlinear example.

Several directions for future work can be pursued. Establishing convergence properties of LMI-QLi, and conditions under which its fixed point is a stationary point of~\eqref{eq:quadratic_Qfunction_cost}, are of interest. For LMI-QL, sufficient conditions under which the convex relaxation is exact, and how they relate to the value of $\lambda$ at which the gap closes, are a natural next step. The connection between the Bellman residual and the value-function error from Section~\ref{sec:method} is idealized. A finite-sample version that holds for the estimated residual, under a persistence-of-excitation assumption, would bound the distance to the optimal policy. A further direction is output feedback, where the full state is not measured.


\bibliographystyle{IEEEtran.bst}

\bibliography{library}

@book{Sutton2018,
  author = {Sutton, Richard S. and Barto, Andrew G.},
  edition = {2},
  publisher = {MIT Press},
  title = {{Reinforcement Learning: An Introduction}},
  year = {2018}
}

@inproceedings{Busoniu2011,
  author = {Bu{\c{s}}oniu, Lucian and Ernst, Damien and {De Schutter}, Bart and Babu{\v{s}}ka, Robert},
  booktitle = {2011 IEEE Symposium on Adaptive Dynamic Programming and Reinforcement Learning (ADPRL)},
  doi = {10.1109/ADPRL.2011.5967353},
  pages = {1--8},
  title = {{Approximate reinforcement learning: An overview}},
  year = {2011}
}

@article{Busoniu2018,
  author = {Bu{\c{s}}oniu, Lucian and de Bruin, Tim and Toli{\'{c}}, Domagoj and Kober, Jens and Palunko, Ivana},
  doi = {10.1016/j.arcontrol.2018.09.005},
  journal = {Annual Reviews in Control},
  pages = {8--28},
  title = {{Reinforcement learning for control: Performance, stability, and deep approximators}},
  volume = {46},
  year = {2018}
}

@article{Xu2014,
  author = {Xu, Xin and Zuo, Lei and Huang, Zhenhua},
  doi = {10.1016/J.INS.2013.08.037},
  journal = {Information Sciences},
  month = {mar},
  pages = {1--31},
  publisher = {Elsevier},
  title = {{Reinforcement learning algorithms with function approximation: Recent advances and applications}},
  volume = {261},
  year = {2014}
}

@article{Bradtke1996,
  author = {Bradtke, Steven J. and Barto, Andrew G.},
  doi = {10.1007/BF00114723},
  journal = {Machine Learning},
  number = {1-3},
  pages = {33--57},
  title = {{Linear Least-Squares algorithms for temporal difference learning}},
  volume = {22},
  year = {1996}
}

@article{Lagoudakis2004,
  author = {Lagoudakis, Michail G. and Parr, Ronald},
  doi = {10.1162/1532443041827907},
  journal = {Journal of Machine Learning Research},
  number = {6},
  pages = {1107--1149},
  title = {{Least-squares policy iteration}},
  volume = {4},
  year = {2003}
}

@incollection{Lagoudakis2014,
  address = {Boston, MA},
  author = {Lagoudakis, Michail G.},
  booktitle = {Encyclopedia of Machine Learning and Data Mining},
  doi = {10.1007/978-1-4899-7502-7_473-1},
  pages = {1--9},
  publisher = {Springer US},
  title = {{Least-Squares Reinforcement Learning Methods}},
  year = {2014}
}

@article{Ernst2005,
  author = {Ernst, Damien and Geurts, Pierre and Wehenkel, Louis},
  journal = {Journal of Machine Learning Research},
  pages = {503--556},
  title = {{Tree-based batch mode reinforcement learning}},
  volume = {6},
  year = {2005}
}

@inproceedings{Riedmiller2005,
  address = {Berlin, Heidelberg},
  author = {Riedmiller, Martin},
  booktitle = {Machine Learning: ECML 2005},
  doi = {10.1007/11564096_32},
  pages = {317--328},
  publisher = {Springer},
  series = {Lecture Notes in Computer Science},
  title = {{Neural Fitted Q Iteration -- First Experiences with a Data Efficient Neural Reinforcement Learning Method}},
  volume = {3720},
  year = {2005}
}

@article{Mnih2015,
  author = {Mnih, Volodymyr and Kavukcuoglu, Koray and Silver, David and Rusu, Andrei A. and Veness, Joel and Bellemare, Marc G. and Graves, Alex and Riedmiller, Martin and Fidjeland, Andreas K. and Ostrovski, Georg and Petersen, Stig and Beattie, Charles and Sadik, Amir and Antonoglou, Ioannis and King, Helen and Kumaran, Dharshan and Wierstra, Daan and Legg, Shane and Hassabis, Demis},
  doi = {10.1038/nature14236},
  journal = {Nature},
  month = {feb},
  number = {7540},
  pages = {529--533},
  publisher = {Nature Publishing Group},
  title = {{Human-level control through deep reinforcement learning}},
  volume = {518},
  year = {2015}
}

@article{Schulman2017,
  author = {Schulman, John and Wolski, Filip and Dhariwal, Prafulla and Radford, Alec and Klimov, Oleg},
  journal = {arXiv preprint arXiv:1707.06347},
  month = {jul},
  title = {{Proximal Policy Optimization Algorithms}},
  year = {2017}
}

@inproceedings{Abdolmaleki2018,
  author = {Abdolmaleki, Abbas and Springenberg, Jost Tobias and Tassa, Yuval and Munos, Remi and Heess, Nicolas and Riedmiller, Martin},
  booktitle = {International Conference on Learning Representations (ICLR)},
  title = {{Maximum a Posteriori Policy Optimisation}},
  year = {2018}
}

@book{Bertsekas1996,
  address = {Belmont, Massachusetts},
  author = {Bertsekas, Dimitri and Tsitsiklis, John N},
  publisher = {Athena Scientific},
  title = {{Neuro-dynamic programming}},
  year = {1996}
}

@incollection{Baird1995,
  author = {Baird, Leemon},
  booktitle = {Machine Learning Proceedings 1995},
  doi = {10.1016/B978-1-55860-377-6.50013-X},
  pages = {30--37},
  publisher = {Elsevier},
  title = {{Residual Algorithms: Reinforcement Learning with Function Approximation}},
  year = {1995}
}

@article{Sanathanan1963,
  author = {Sanathanan, C. and Koerner, J.},
  doi = {10.1109/TAC.1963.1105517},
  journal = {IEEE Transactions on Automatic Control},
  month = {jan},
  number = {1},
  pages = {56--58},
  title = {{Transfer function synthesis as a ratio of two complex polynomials}},
  volume = {8},
  year = {1963}
}

@article{Recht2010,
  author = {Recht, Benjamin and Fazel, Maryam and Parrilo, Pablo A.},
  doi = {10.1137/070697835},
  journal = {SIAM Review},
  month = {jan},
  number = {3},
  pages = {471--501},
  title = {{Guaranteed Minimum-Rank Solutions of Linear Matrix Equations via Nuclear Norm Minimization}},
  volume = {52},
  year = {2010}
}

@article{ElGhaoui1997,
  author = {{El Ghaoui}, L. and Oustry, F. and AitRami, M.},
  doi = {10.1109/9.618250},
  journal = {IEEE Transactions on Automatic Control},
  number = {8},
  pages = {1171--1176},
  title = {{A cone complementarity linearization algorithm for static output-feedback and related problems}},
  volume = {42},
  year = {1997}
}

@article{Freeman1996,
  author = {Freeman, R. A. and Kokotovic, P. V.},
  doi = {10.1137/S0363012993258732},
  journal = {SIAM Journal on Control and Optimization},
  month = {jul},
  number = {4},
  pages = {1365--1391},
  title = {{Inverse Optimality in Robust Stabilization}},
  volume = {34},
  year = {1996}
}

\end{document}